\documentclass[twocolumn,showpacs,preprintnumbers,amsmath,amssymb]{revtex4}
\usepackage{amsfonts}
\usepackage{graphicx}
\usepackage{dcolumn}
\usepackage{bm}

\begin{document}
\preprint{quant-phys}
\title[Exact solution]{Exact solution of qubit decoherence models by a transfer matrix method}
\author{Diu Nghiem, Robert Joynt}
\affiliation{Dept. of Physics, University of Wisconsin-Madison, 1150 Univ. Ave., Madison, 53706}

\begin{abstract}
We present a new method for the solution of the behavior of an ensemble of
qubits in a random time-dependent external field. \ The forward evolution in
time is governed by a transfer matrix. \ The elements of this matrix determine
the various decoherence times. \ The method provides an exact solution in
cases where the noise is piecewise constant in time. \ We show that it
applies, for example, to a realistic model of decoherence of electron spins in
semiconductors. \ Results are obtained for the non-perturbative regimes of the
models, and we see a transition from weak relaxation to overdamped behavior as
a function of noise anisotropy. \ 

\end{abstract}
\volumeyear{year}
\volumenumber{number}
\issuenumber{number}
\eid{identifier}
\date[Date text]{date}
\received[Received text]{date}

\revised[Revised text]{date}

\accepted[Accepted text]{date}

\published[Published text]{date}

\maketitle


\section{Introduction}

In the last decade, advances in fabrication and experimentation have made it
possible to observe and control quantum effects in individual systems that
could previously be observed only in the aggregate. \ One important condition
for these advances has been to minimize, or at least to manage, decoherence.
\ Quantum systems are inevitably subject to random influences that destroy
coherence, and classical behavior takes over. \ \ Only if this can be avoided
can we hope to advance the technology of quantum information processing, and
theoretical understanding can help to achieve this. \ Theoretical calculations
generally fall into three categories: eactly soluble examples, approximation
methods, and numerical computations. \ The theory of quantum decoherence has
been notable for its richness in the last two of these three, but exactly
soluble models have been few and far between, though some exceptions may be
noted \cite{unruh},\cite{palma},\cite{khaetskii}. \ This paper goes some way
towards changing this situation. \ 

Most treatments of decoherence begin with two coupled quantum objects: the
system to be observed, and the unmeasured environment to which the system is
coupled. \ The system eventually loses its coherence due to that coupling.
\ The most natural way to treat the problem is \ to formulate a master
equation for the reduced density matrix, which incorporates the
system-environment coupling, and also some aspects of the environment's
dynamics. \ This approach can in principle treat the "back-action" of the
system on the environment, though this is often difficult in practice. \ The
classic example is the Caldeira-Leggett model \cite{caldeira} \ 

An alternative formulation is to treat the environment as a fixed source of
random noise. \ In this approach it is not possible to take into account
back-action. \ However, there are many physical situations in which
back-action is not important. \ Treating the environment as fixed but random
is then usually more convenient. \ This approach will be used in this paper.

In quantum computation, a large array of qubits - two-level quantum systems -
needs to maintain coherence. \ A necessary requirement for this is to maintain
quantum coherence of a single qubit, and most discussions begin at this level.
\ The restriction to a two-state quantum system allows simplifications in the
formalism. \ The purpose of this paper is to show that the decoherence of a
two-state system subject to random noise can, under certain conditions, be
treated by a powerful transfer matrix method analogous to that of statistical
mechanics, allowing us to solve a number of cases exactly. \ Furthermore,
these solutions apply to the non-perturbative situation in which the coupling
to the environment is not smaller than the level separation of the qubit.
\ This difficult regime is of great importance since some quantum computation
schemes, most notably holonomic quantum computing, propose to operate in this
parameter range \cite{zanardi}.

In the next section we introduce and solve a class of white-noise-type models
that can be solved by a relatively straightforward application of the method.
\ We discuss a simple example, and compare the exact results to perturbation
theory. \ The new results in the non-perturbative regime are also explored.
\ \ In the third section we extend the method to Markovian models that can
also be solved, this time by a slightly more intricate form of the transfer
matrix. \ In the final section we discuss the possible extensions and ultimate
limitations of the method.

\section{White-noise-like models\bigskip}

\subsection{Definition}

Our two-level system will be described by the Hamiltonian%
\begin{equation}
H=-\vec{B}_{0}\sigma_{z}-\vec{b}(t)\cdot\vec{\sigma}, \label{eq:hamiltonian}%
\end{equation}
so that the energy separation is $2B_{0}$ and $\vec{b}(t)$ is a random
function. \ The system is prepared at time $t=0$ with a $2\times2$ density
matrix $\rho(t=0)=\rho_{0}.$ \ The measurable quantities are \
\begin{equation}
\overline{\vec{M}(t=t_{f})}=\int\mathcal{P}[\vec{b}(t)]\mathcal{D}[\vec
{b}(t)]~Tr~\left[  \rho_{0}\vec{\sigma}(t)\right]  . \label{eq:rho}%
\end{equation}
\ Here $\sigma_{x},\sigma_{y},$ and $\sigma_{z}$ are the Pauli matrices, which
together with the unit matrix represent a complete set of observables in the
qubit space. \ We work in the Heisenberg representation, so that $\vec{\sigma
}(t)$ satisfies $i~d\vec{\sigma}/dt=\left[  \vec{\sigma},H\right]  $.
\ $\int\mathcal{P}[\vec{b}(t)]\mathcal{D}[\vec{b}(t)]$ indicates a sum over
all possible functions $\vec{b}(t)$ on the interval $t\in\lbrack0,t_{f}],$
with the appropriate probability $\mathcal{P}[\vec{b}(t)].$ \ The
specification of $\mathcal{P}[\vec{b}(t)]$ defines a model in the general
category of `qubits subject to random time-dependent forces. \ 

Our notation is of course motivated by spin qubits. \ However, it should be
obvious that the method and all results apply to any two-level system. \ In
particular, \ $\vec{B}_{0}$ and $\vec{b}$ need not be magnetic fields.

We now consider the following form of $\mathcal{P}[\vec{b}(t)].$ \ Let
$\vec{b}(t)$ be piecewise constant with discontinuous jumps at regular
intervals of length $\tau.$ \ Furthermore, let $\vec{b}(t)$ be completely
independent between time intervals, i.e., $\vec{b}(t)=\vec{b}_{1}$ and
$\vec{B}(t)=B_{0}\widehat{z}+\vec{b}_{1}=\vec{B}_{1}$ for $0<t<\tau,$ $\vec
{b}(t)=\vec{b}_{2}$ and $\vec{B}(t)=B_{0}\widehat{z}+\vec{b}_{2}$ for
$\tau<t<2\tau,$ etc. , with no correlation between the different $\vec{b}%
_{i}.$ \ \ Finally, each $\vec{b}_{i}$ has the same probability distribution
$P(\vec{b})$. \ Let there be $m$ time intervals so that the final time when a
measurement is performed is $t_{f}=m\tau.$ \ For this model, then, we have
\begin{align*}
\int&\mathcal{P}[\vec{b}(t)]\mathcal{D}[\vec{b}(t)]F\left[ \vec{b}(t)\right]\\
& =\int d^{3}b_{1}P(\vec{b}_{1})\cdot\cdot\cdot\ \int d^{3}b_{m}P(\vec{b}%
_{m})F\left[  \vec{b}(t)\right] \\
&  =\overline{F\left[  \vec{b}(t)\right]  }%
\end{align*}
for the average of any functional $F\left[  \vec{b}(t)\right]  $. $
P(\vec{b})\geq0$ and $\int d^{3}b~P(\vec{b})=1.$ \ Any model with the
piecewise constant $\vec{b}$ on equally-spaced intervals and the product form
for the probability is exactly soluble, as we now show. \ We shall also give a
physical example below, in order to indicate that the model is not artificial.

\subsection{General solution}

Defining
\[
H_{i}=-\vec{B}_{i}\cdot\vec{\sigma}=-\vec{B}_{0}\sigma_{z}-\vec{b}_{i}%
\cdot\vec{\sigma}%
\]
and%
\[
U_{i}=e^{-iH_{i}\tau},
\]
the solution for $\overline{\vec{\sigma}(t_{f})}$ is%

\begin{equation}
\overline{\vec{\sigma}(t_{f})}=\overline{\vec{\sigma}(m\tau)}=\overline
{U_{m}^{+}\cdot\cdot\cdot U_{2}^{+}U_{1}^{+}\vec{\sigma}(0)U_{1}U_{2}%
\cdot\cdot\cdot U_{m}}.
\end{equation}
Once we have $\overline{\vec{\sigma}(t_{f})}$ we can apply the initial
conditions and obtain results for any observable according to Eq.
\ref{eq:rho}. \ Because of the statistical independence of the values 
of $\vec{b}$ at different intervals,
the averaging can be done for each correlation time individually, so
$\vec{\sigma}_{ij}$ transforms in the time $\tau$ into%
\begin{widetext}
\begin{align*}
\overline{\vec{\sigma}_{ij}(\tau)}  &  =\int d^{3}b_{1}P(\vec{b}_{1}%
)\overline{\left(  U_{1}^{+}\right)  _{ik}\vec{\sigma}_{kl}(0)\left(
U_{1}\right)  _{lj}}\\
&  =\overline{\left[  \exp\left(  iH_{1}\tau\right)  \right]  _{ik}\left[
\exp\left(  -i\vec{H}_{1}\tau\right)  \right]  _{lj}}\vec{\sigma}_{kl}(0)\\
&  =\overline{\left[  E\cos(B_{1}\tau)-i\widehat{B}_{1}\cdot\vec{\sigma}%
\sin(B_{1}\tau)\right]  _{ik}\left[  E\cos(B_{1}\tau)+i\widehat{B}_{1}%
\cdot\vec{\sigma}\sin(B_{1}\tau)\right]  _{lj}}\vec{\sigma}_{kl}(0),
\end{align*}
\end{widetext}
and summation over repeated indices is implied. \ The identity matrix
$E$ always transforms into
itself: $E(\tau)=E.$ \ We have used the identity $\exp(i\vec{X}\cdot
\vec{\sigma}t)=\cos Xt+i\widehat{X}\cdot\vec{\sigma}\sin Xt,$ valid for any
vector $\vec{X}.$ \ $\widehat{X}=\vec{X}/\left\vert \vec{X}\right\vert
=\vec{X}/X.$ \ The key to the method is to compute and then iterate the
superoperator $\overline{\left[  U_{1}^{+}(\tau)\right]  _{ik}|\left[
U_{1}(\tau)\right]  _{lj}}.$\ \ Performing the average and writing the results
in matrix form, we find%
\begin{equation}
\overline{\vec{\sigma}(\tau)}=I_{0}\vec{\sigma}(0)+\sum_{ij}I_{ij}\sigma
_{i}\vec{\sigma}(0)\sigma_{j}+i\sum_{i}I_{i}\left[  \vec{\sigma}(0)\sigma
_{i}-\sigma_{i}\vec{\sigma}(0)\right]  ,
\end{equation}
where the integrals $I_{0},~I_{i}$ and $I_{ij}$ are defined as%
\begin{align}
I_{0}  &  =\int P(\vec{b})\cos^{2}(B\tau)d^{3}b\label{eq:integrals}\nonumber\\
I_{i}  &  =\int P(\vec{b})\widehat{B}_{i}\sin(B\tau)\cos\left(  B\tau\right)
d^{3}b\\
I_{ij}  &  =\int P(\vec{b})\widehat{B}_{i}\widehat{B}_{j}\sin^{2}(B\tau
)d^{3}b.\nonumber
\end{align}
Here $\vec{B}=\vec{B}_{0}+\vec{b},~\widehat{B}_{x}=b_{x}/B,$ $\widehat{B}%
_{y}=b_{y}/B,$ and $\widehat{B}_{z}=\left(  b_{z}+B_{0}\right)  /B.$ \ (Note
that $\vec{B}_{0}$ may be defined so that $I_{x}=I_{y}=0.$ \ In order to
exhibit the symmetry of the following expressions, however, we shall retain
these quantitities.) \ Using the properties of the Pauli matrices to simplify
the expressions, we find the three operators$:$%
\begin{widetext}
\begin{align}
\overline{\sigma_{x}(\tau)}&=\left(  I_{0}+I_{xx}-I_{yy}-I_{zz}\right)
\sigma_{x}(0)+2\left(  I_{xy}+I_{z}\right)  \sigma_{y}(0)+2\left(
I_{xz}-I_{y}\right)  \sigma_{z}(0)\nonumber\\
\overline{\sigma_{y}(\tau)}&=\left(  I_{0}-I_{xx}+I_{yy}-I_{zz}\right)
\sigma_{y}(0)+2\left(  I_{xy}-I_{z}\right)  \sigma_{x}(0)+2\left(
I_{yz}+I_{x}\right)  \sigma_{z}(0)\\
\overline{\sigma_{z}(\tau)}&=\left(  I_{0}-I_{xx}-I_{yy}+I_{zz}\right)
\sigma_{z}(0)+2\left(  I_{xz}+I_{y}\right)  \sigma_{x}(0)+2\left(
I_{yz}-I_{x}\right)  \sigma_{y}(0).\nonumber
\end{align}
\end{widetext}
We may write this in a compact fashion by defining the transfer matrix $T$ by%
\[%
\begin{pmatrix}
\overline{\sigma_{x}(\tau)}\\
\overline{\sigma_{y}(\tau)}\\
\overline{\sigma_{z}(\tau)}%
\end{pmatrix}
=%
\begin{pmatrix}
T_{xx} & T_{xy} & T_{xz}\\
T_{yx} & T_{yy} & T_{yz}\\
T_{zx} & T_{zy} & T_{zz}%
\end{pmatrix}%
\begin{pmatrix}
\sigma_{x}(0)\\
\sigma_{y}(0)\\
\sigma_{z}(0)
\end{pmatrix}
,
\]
or%
\[
\vec{\sigma}(\tau)=T~\vec{\sigma}(0),
\]
for short - it is understood that in expressions of this kind $\vec{\sigma}$
is a column vector. \ The detailed expressions for the matrix elements are: \
\begin{align}
T_{xx}  &  =I_{0}+I_{xx}-I_{yy}-I_{zz}\nonumber\\
T_{yy}  &  =I_{0}-I_{xx}+I_{yy}-I_{zz}\nonumber\\
T_{zz}  &  =I_{0}-I_{xx}-I_{yy}+I_{zz}\nonumber\\
T_{xy}  &  =2I_{xy}+2I_{z}\nonumber\\
T_{yx}  &  =2I_{xy}-2I_{z}\label{eq:matrix}\\
T_{xz}  &  =2I_{xz}-2I_{y}\nonumber\\
T_{zx}  &  =2I_{xz}+2I_{y}\nonumber\\
T_{yz}  &  =2I_{yz}+2I_{x}\nonumber\\
T_{zy}  &  =2I_{yz}-2I_{x}.\nonumber
\end{align}
$T$ is an example of an exactly computable quantum dynamical map. \ For a
definition and general discussion of such maps, see Ref. \cite{breuer}. \ At
$t_{f},$ we find%
\begin{align}
\overline{\vec{\sigma}(t_{f})}  &  =\overline{\vec{\sigma}(m\tau)}\nonumber\\
&  =\int d^{3}b_{m}P(\vec{b}_{m})U_{m}^{+}\cdot\cdot\cdot\int d^{3}b_{2}%
P(\vec{b}_{2}) \times \\
& \,\,\,\,\left[  U_{2}^{+}\left(  \int d^{3}b_{1}P(\vec{b}_{1})U_{1}%
^{+}\vec{\sigma}(0)U_{1}\right)  U_{2}\right]  \cdot\cdot\cdot U_{m},\nonumber
\end{align}
so this process can be iterated, and we have%
\[
\vec{\sigma}(t_{f})=\vec{\sigma}(m\tau)=T^{m}~\vec{\sigma}(0).
\]
$T$ is real but generally not symmetric. \ It operates in the complex
three-dimensional vector space of superoperators$.$ \ Each vector
$(a_{x},a_{y},a_{z})$ in the space is associated with the operator $\vec
{a}\cdot\vec{\sigma}.$ \ 

Diagonalizing $T,$ we find
\[
D=RTR^{-1}%
\]
where $D=diag(d_{1},d_{2},d_{3})$ and the rows of $R$ are the eigenvectors of
$T.$ \ Since $T$ is real, the eigenvalues are the roots of a cubic equation
with real coefficients, so at least one of the $d_{i}$ is real and the other
two must either be real, or they must be complex conjugates. \ Iterating, we
have
\begin{align*}
\vec{\sigma}(t_{f})  &  =\vec{\sigma}(m\tau)\\
&  =T^{m}~\vec{\sigma}(0)\\
&  =\left(  R^{-1}DR\right)  ^{m}\vec{\sigma}(0)\\
&  =R^{-1}D^{m}R~\vec{\sigma}(0),
\end{align*}
with%
\[
D^{m}=%
\begin{pmatrix}
\left(  d_{1}\right)  ^{m} & 0 & 0\\
0 & \left(  d_{2}\right)  ^{m} & 0\\
0 & 0 & \left(  d_{3}\right)  ^{m}%
\end{pmatrix}
.
\]
Recalling that $m=t_{f}/\tau$ and using index notation, the solution becomes%
\begin{equation}
\sigma_{i}(t_{f})=R_{ij}^{-1}\left(  d_{j}\right)  ^{t_{f}/\tau}R_{jk}%
~\sigma_{k}(0). \label{eq:solution}%
\end{equation}
This equation, together with Eqs. \ref{eq:integrals}, \ref{eq:matrix}, and
\ref{eq:rho} constitute an exact solution for all $P(\vec{b}).$ \ The
magnitudes of the eigenvalues $d_{j}$ determine the relaxation times through
the relation%
\begin{equation}
\frac{1}{T_{j}}=-\frac{1}{\tau}\ln\left\vert d_{j}\right\vert ,
\end{equation}
while the transformation matrices $R$ contain the information that allows an
experimenter to prepare and measure a particular relaxation time. \ 

\subsection{Realistic Example}

A relatively simple example of current interest concerns the spin 
dynamics of an electron in a two-dimensional system such as an
appropriately constructed semiconductor heterostructure. \
Let the system be subject to an electric field $\vec{E}=E\widehat{z}$ 
and an applied uniform time-independent magnetic field 
$B_{0}\widehat{z}$ perpendicular to the conducting layer. \ Due to
spin-orbit couping, there is also an effective magnetic field acting on the
spin, according to $\vec{b}=\alpha\vec{p}\times\vec{E},$ where the electron
momentum $\vec{p}$ lies in the $x-y$ plane. \ $\alpha$ is a
constant and $\left\vert \overrightarrow{p}\right\vert =p_{F},$ the Fermi
momentum. \ The electron scatters, causing $\vec{p}$ and therefore $\vec{b}$
to be a random function of time. \ The magnitudes $\left\vert \vec
{p}\right\vert $ and therefore $\left\vert \vec{b}\right\vert =b_{0}$ are
fixed, however. \ If we take the scattering events to occur at equally-spaced
intervals of length $\tau,$ now understood as a time related to the momentum
scattering time $\tau_{p}$, then this situation falls into the class of models
considered in this paper. \ The assumption of \ equally-spaced intervals is of
course an approximation - the actual length of the intervals is random, and is
governed by a Poisson distribution. \ The effective magnetic field is called
the Rashba field \cite{bychkov}, and the spin relaxation mechanism is known as
the DP or D'yakonov-Perel' mechanism \cite{d'yakonov}. \ 

Hence the Hamiltonian for the spin degree of freedom is precisely of the form
shown in Eq.\ref{eq:hamiltonian}, and in this case $\vec{b}$ lies in the
$x-y$ plane. \ The average over $\vec{b}$ is performed by
settting $\vec{b}=b_{0}(\cos\phi,\sin\phi,0)$ and averaging uniformly over
$\phi.$ \ $\vec{B}=B_{0}\widehat{z}+\vec{b}.$ \ In the physical situation of
interest, the environmental correlation time $\tau$ is usually much less than
the spin relaxation times and also much less than $1/B_{0}.$ \ Now set
$I_{x}=I_{y}=0,$ and note that symmetry in the $x-y$ plane also gives
$I_{xy}=I_{yx}=I_{xz}=I_{zx}=I_{yz}=I_{zy}=0.$ \ For this particular case
$I_{zz}=0$ since $b_{z}=0,$ but it is interesting to keep this quantity for
the time being. \ The transfer matrix simplifies considerably:
\begin{align}
T_{xx}  &  =I_{0}+I_{xx}-I_{yy}-I_{zz}\nonumber\\
T_{yy}  &  =I_{0}-I_{xx}+I_{yy}-I_{zz}\nonumber\\
T_{zz}  &  =I_{0}-I_{xx}-I_{yy}+I_{zz}\\
T_{xy}  &  =-T_{yx}=2I_{z}\nonumber\\
T_{xz}  &  =I_{zx}=I_{yz}=I_{zy}=0\nonumber
\end{align}
The eigenvalues of $T$ are $d_{z}=T_{zz}=I_{0}-I_{xx}-I_{yy}+I_{zz},$ and
$d_{x,y}=I_{0}-I_{zz}\pm i\sqrt{4I_{z}^{2}-\left(  I_{xx}-I_{yy}\right)  ^{2}%
}.$ \ \ In the Appendix we show that these eigenvalues satisfy $\left\vert
d_{3}\right\vert \leq1$ and $\left\vert d_{\pm}\right\vert \leq1,$ which
guarantees that all solutions are exponentially decaying.

The connection to decoherence times is obtained by considering the limit where
$t_{f}>>1/B_{0}>>\tau$ and $t_{f}>>1/\left\vert \vec{b}\right\vert >>\tau
$.\ Expanding the integrals for small $\tau,$ we find%
\begin{equation}
d_{z} \approx  1-2\tau^{2}\int P(\vec{b})\left(  b_{x}^{2}+b_{y}^{2}\right)
d^{3}b=1-2\tau^{2}\left(  \overline{b_{x}^{2}}+\overline{b_{y}^{2}}\right)
\label{eq:dz} 
\end{equation}
\begin{align}
d_{x,y}  \approx & 1-\tau^{2}\int P(\vec{b})\left(  B_{0}^{2}+b^{2}\right)
d^{3}b +  \nonumber\\
\, & \,-\tau^{2}\int P(\vec{b})B_{0}^{2}~d^{3}b\pm2i\sqrt{\tau^{2}B_{0}%
^{2}-\left(  I_{xx}-I_{yy}\right)  ^{2}/4}\nonumber \\
  = &1-2\tau^{2}B_{0}^{2}-\tau^{2}\left(  \overline{b_{x}^{2}}+\overline
{b_{y}^{2}}+2\overline{b_{z}^{2}}\right) + \nonumber\\
\,&\, \pm2i\sqrt{\tau^{2}B_{0}^{2}%
-\frac{1}{4}\tau^{4}\left(  \overline{b_{x}^{2}}-\overline{b_{y}^{2}}\right)
^{2}~}\label{eq:dxy}%
\end{align}
The result for $d_{z}$ is independent of the ratio $\left\vert \vec
{b}\right\vert /B_{0},$ but the $d_{x,y}$ are sensitive to it. \ Most
experiments are done in the regime where $\left\vert \vec{b}\right\vert
<<B_{0},$ and we consider this perturbative case first:%
\begin{align}
\left\vert d_{x,y}\right\vert & =\left\vert 1-2\tau^{2}B_{0}^{2}-\tau
^{2}\left( \overline{b_x^2}+\overline{b_y^2}+2\overline{b_z^2}\right)
\pm 2i\tau B_0+O(\tau^4) \right\vert\nonumber \\
&  \approx1-\tau^{2}\left(  \overline{b_{x}^{2}}+\overline{b_{y}^{2}%
}+2\overline{b_{z}^{2}}\right)
\end{align}
If the system is prepared in an energy eigenstate, (in spin language, along
the z-direction), then it remains along that direction since $T_{xz}%
=T_{yz}=0,$ and we find
\begin{align}
\left\langle \sigma_{z}(t_{f})\right\rangle  &  =\left(  d_{3}\right)
^{t_{f}/\tau}\sigma_{z}(0)\nonumber\\
&  =\left[  1-2\tau^{2}\left(  \overline{b_{x}^{2}}+\overline{b_{y}^{2}%
}\right)  \right]  ^{t_{f}/\tau}\left\langle \sigma_{z}(0)\right\rangle .
\end{align}
This is finite in the limit of small $1/b$ and $\tau$ if the combination
$\tau\left(  \overline{b_{x}^{2}}+\overline{b_{y}^{2}}\right)  $ approaches a
finite limit, and we have%
\begin{equation}
\left\langle \sigma_{z}(t_{f})\right\rangle \rightarrow\exp\left(
-t_{f}/T_{1}\right)  \left\langle \sigma_{z}(0)\right\rangle ,
\end{equation}
with the energy relaxation time $T_{1}$ given by%
\begin{equation}
\frac{1}{T_{1}}=2\tau\left(  \overline{b_{x}^{2}}+\overline{b_{y}^{2}}\right)
=2b_{0}^{2}\tau. \label{eq:t1}%
\end{equation}
If the system is prepared in an equal superposition of the energy eigenstates
(in spin language, in the x-y plane), then there are oscillations of
$\sigma_{x}$ and $\sigma_{y}$ owing to the fact that the $d_{x,y}$ are not
real, at an angular frequency given by $\phi/\tau,$ where $\phi=\left\vert
\tan^{-1}\left(  \operatorname{Im}d_{+}/\operatorname{Re}d_{+}\right)
\right\vert .$ \ There is also exponential decay given by%
\begin{equation}
\left\vert \left\langle \sigma_{x,y}(t_{f})\right\rangle \right\vert
\rightarrow\exp\left(  -t_{f}/T_{2}\right)  \left\vert \left\langle
\sigma_{x,y}(0)\right\rangle \right\vert
\end{equation}
with the phase relaxation time $T_{2}$ given by%
\begin{equation}
\frac{1}{T_{2}}=\tau\left(  \overline{b_{x}^{2}}+\overline{b_{y}^{2}%
}+2\overline{b_{z}^{2}}\right)  =b_{0}^{2}\tau. \label{eq:t2}%
\end{equation}
Note that these results satisfy the constraint that in the isotropic
($B_{0}\rightarrow0$) limit $\overline{b_{x}^{2}}=\overline{b_{y}^{2}%
}=\overline{b_{z}^{2}}$ we must have $T_{1}=T_{2},$ since the distinction
between longitudinal and transverse relaxation ceases to have any meaning.
\ Incidentally, the perturbation theory results for the same model except that
the time intervals follow Poisson statistics are: $1/T_{1}=$ $4\tau_{p}\left(
\overline{b_{x}^{2}}+\overline{b_{y}^{2}}\right)  $ and $1/T_{2}=2\tau
_{p}\left(  \overline{b_{x}^{2}}+\overline{b_{y}^{2}}+2\overline{b_{z}^{2}%
}\right)  $~\cite{tahan}$.$ \ Thus the previous results can be translated over
with the identification $\tau=2\tau_{p}.$

The rotation matrix $R$ is block-diagonal for this simplified case:
$R_{xz}=R_{zx}=R_{yz}=R_{zy}=0.$ \ This means that preparation and measurement
of the spin in the $z-$direction will give $T_{1}$ and preparation and
measurement of the spin in any direction in the $x-y$ plane will give $T_{2}.$
\ Preparation in any other direction will result in independent relaxations of
the $z-$ component and the magnitude of the projection of the spin onto the
$x-y$ plane. \ There is no cross-talk between the two relaxations.

\subsection{Comparison with perturbation theory}

The usual treatment of problems in this class is by perturbation theory, often
going under the name of Redfield theory in this context. \ The theory has been
extensively worked out because of its applications in nuclear magnetic
resonance. \ The condition for the validity of the perturbative approach is
$B_{0}>>b(t)$ at all times. \ A detailed treatment is given by Slichter
\cite{slichter}. \ The results for relaxation times are:%
\begin{align}
\frac{1}{T_{1}}  &  =k_{xx}(\omega_{0})+k_{yy}(\omega_{0})\label{eq:kzz}\\
\frac{1}{T_{2}}  &  =\frac{1}{2T_{1}}+k_{zz}(0),
\end{align}
with $\omega_{0}=2B_{0}$ and%
\begin{equation}
k_{ii}(\omega)=2\int_{-\infty}^{\infty}ds~e^{-i\omega s}\overline
{b_{i}(t)b_{i}(t+s)}. \label{eq:kii}%
\end{equation}
To compute $k_{ii}$ in the example of the previous subsection, we note that
$k_{xx}=k_{yy}$ and $k_{zz}=0$ and that we\ must average $t$ uniformly over
the interval $\left(  0,\tau\right)  $ and average $\phi$ uniformly over the
interval $\left(  0,2\pi\right)  .$ \ For the piecewise constant $b(t),$ we
find
\begin{align}
\overline{b_{i}(t)b_{i}(t+s)}=&b^2_{0}\overline{\cos\left[  \phi(t)\right]
~\cos\left[  \phi(t+s)\right]  } \nonumber \\
\,=&\left\{
\begin{array}{ccc}%
0,& \text{if}& s<-t\\
b_{0}^{2}/2,& \text{if} &-t<s<\tau-t\\
0,&\text{if}&  s>\tau-t.
\end{array}
\right.
\end{align}
This result can be substituted into Eq. \ref{eq:kii}. In the limit that
$B_{0}\tau<<1$, $k_{xx}(\omega_{0})\rightarrow b_{0}^{2}\tau,$ and similarly
for $k_{yy}(\omega_{0}).$ \ Thus perturbation theory applied to our model
yields\ $1/T_{1}=2b_{0}^{2}\tau$ and $1/T_{2}=1/2T_{1},$ in agreement
with\ Eqs. \ref{eq:t1} and \ref{eq:t2}. \ Although we have not treated them in
detail, the longitudinal fluctuations described by the function $k_{zz}$ that
produce dephasing also have the correct perturbative limit [cf. Eqs.
\ref{eq:t2} and \ref{eq:kzz}]. \ 

\subsection{Non-perturbative regime}

In this subsection, we wish to investigate the overall behavior of the exact
solution. \ There are four parameters in the model: $\tau,1/B_{0},1/\left\vert
\vec{b}\right\vert ,$ and $\tau_{f}.$ \ The regime $t_{f}<\tau$\ is of no
interest: the spin precesses freely about the random field $\vec{B}_{0}%
+\vec{b}_{1}.$ \ So we focus on the regime $t_{f}>>\tau,$ and the expression
for the observables in Eq. \ref{eq:solution}. \ We have already discussed the
perturbative regime in which $t_{f}>>1/\left\vert \vec{b}\right\vert
>>1/B_{0}>>\tau.$ Our solution is non-perturbative, and remains valid for
arbitrary values of $\left\vert \vec{b}\right\vert /B_{0},$ $\left\vert
\vec{b}\right\vert \tau$ and $B_{0}\tau.$ \ \ However, simple expressions are
available only in relatively symmetric situations, so let us take $\left\vert
\vec{b}\right\vert =b_{0}$ and $I_{xy}=I_{yx}=I_{xz}=I_{zx}=I_{yz}=I_{zy}=0,$
but allow for the possibility that $I_{xx}\neq I_{yy},$ and also that
$b_{z}\neq0$. \ We also assume the probability density $P(\vec{b})$posesses
well-defined moments up to the fourth order in $\vec{b},$ and that the odd
moments vanish. \ We once again have $d_{z}=T_{zz}=I_{0}-I_{xx}-I_{yy}%
+I_{zz}.$

Since we have a closed-form solution, we can easily compute higher-order
expansions. \ To see this, take $\left\vert \vec{b}\right\vert \tau<<1$ and
$B_{0}\tau<<1$ in order to perform the trigonometric integrals. \ Then to
order $\tau^{4},$ we have%
\begin{widetext}
\begin{align}
\frac{1}{T_{1}} = -\frac{1}{\tau}\ln\left\vert d_{z}\right\vert
  = 2\left(\overline{b_{x}^{2}}+\overline{b_{y}^{2}}\right) \tau+\left\{
2\left(  \overline{b_{x}^{2}}+\overline{b_{y}^{2}}\right)  ^{2}-\frac{1}%
{3}\left[  2B_{0}^{2}\left(  \overline{b_{x}^{2}}+\overline{b_{y}^{2}}\right)
+2\overline{b_{x}^{4}}+2\overline{b_{y}^{4}}+2\overline{b_{x}^{2}b_{y}^{2}%
}+\overline{b_{x}^{2}b_{z}^{2}}+\overline{b_{y}^{2}b_{z}^{2}}\right]
\right\}  \tau^{3}%
\end{align}
The strength of the steady field enters the relaxation rate at order $\tau
^{3}.$

For $T_{2}$ there are two possibilities. \ For $I_{z}>\left\vert I_{xx}%
-I_{yy}\right\vert ,$ $d_{x}$ has an imaginary part, $d_{x,y}=I_{0}-I_{zz}\pm
i\sqrt{4I_{z}^{2}-\left(  I_{xx}-I_{yy}\right)  ^{2}}$, $\left\vert
d_{x}\right\vert ^{2}=\left(  I_{0}-I_{zz}\right)  ^{2}+4I_{z}^{2}-\left(
I_{xx}-I_{yy}\right)  ^{2}$ and
\begin{align}
\frac{1}{T_{2}} &  =-\frac{1}{\tau}\ln\left\vert d_{x}\right\vert =-\frac
{1}{\tau}\ln\left\vert d_{y}\right\vert \\
&  =\tau\left(  \overline{b_{x}^{2}}+\overline{b_{y}^{2}}+2\overline{b_{z}%
^{2}}\right)  +\tau^{3}\left[  \left(  \overline{b_{x}^{2}}+\overline
{b_{y}^{2}}+2\overline{b_{z}^{2}}\right)  ^{2}-\frac{1}{3}B_{0}^{2}b_{+}%
^{2}+\frac{1}{3}\left(  \overline{b_{x}^{4}}+\overline{b_{y}^{4}}%
+2\overline{b_{z}^{4}}+2\overline{b_{x}^{2}b_{y}^{2}}+3\overline{b_{+}%
^{2}b_{z}^{2}}\right)  -\frac{1}{2}\left(  \overline{b_{x}^{2}}-\overline
{b_{y}^{2}}\right)  ^{2}\right]\nonumber
\end{align}
\end{widetext}
again good to order $\tau^{4}.$ \ This corresponds to a damped oscillatory
solution. For $I_{z}<\left\vert I_{xx}-I_{yy}\right\vert ,$ $\ $the $d_{x,y}$
are real and the solution is overdamped. \ All three eigenvalues of $T$ are
real, and so there are three relaxation times. \ \ $T_{1}$ remains as above,
and, to order $\tau^{2},$ we find%

\[
\frac{1}{T_{2\pm}}=\overline{\left(  b_{x}^{2}+b_{y}^{2}+2b_{z}^{2}\right)
}\tau\pm\sqrt{\left(  \overline{b_{x}^{2}-b_{y}^{2}}\right)  ^{2}-4B_{0}^{4}%
}\tau.
\]
Here the spin relaxes so fast that it does not have time to precess about the
external field. \ 

Of particular interest is the fact that this abrupt transition from
oscillating to overdamped behavior is driven by the \textit{anisotropy} of the
noise. \ Physically, strong highly anisotropic noise quickly erases the
distinction between the clockwise and counterclowise directions about the
$z$-axis, and the spin does not know which way to precess.

The most interesting issue that is accessible for the exact solution is the
crossover from the weak-noise regime $(\left\vert \vec{b}\right\vert <<B_{0})$
to the strong-noise $(\left\vert \vec{b}\right\vert >>B_{0})$ regime. \ \ \ 

{\begin{figure}[ptb]
\begin{tabular}{l}
{(a)}\\
\includegraphics[scale=0.67, angle=-90]{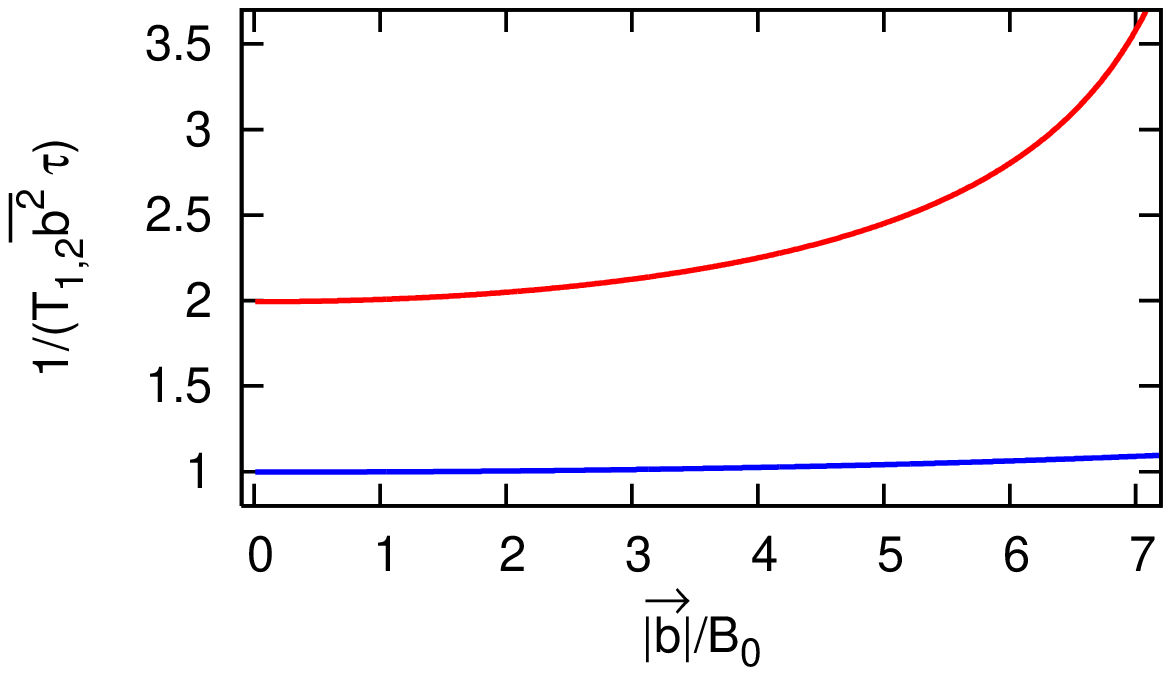}\\ \\
(b)\\
\includegraphics[scale=0.67, angle=-90]{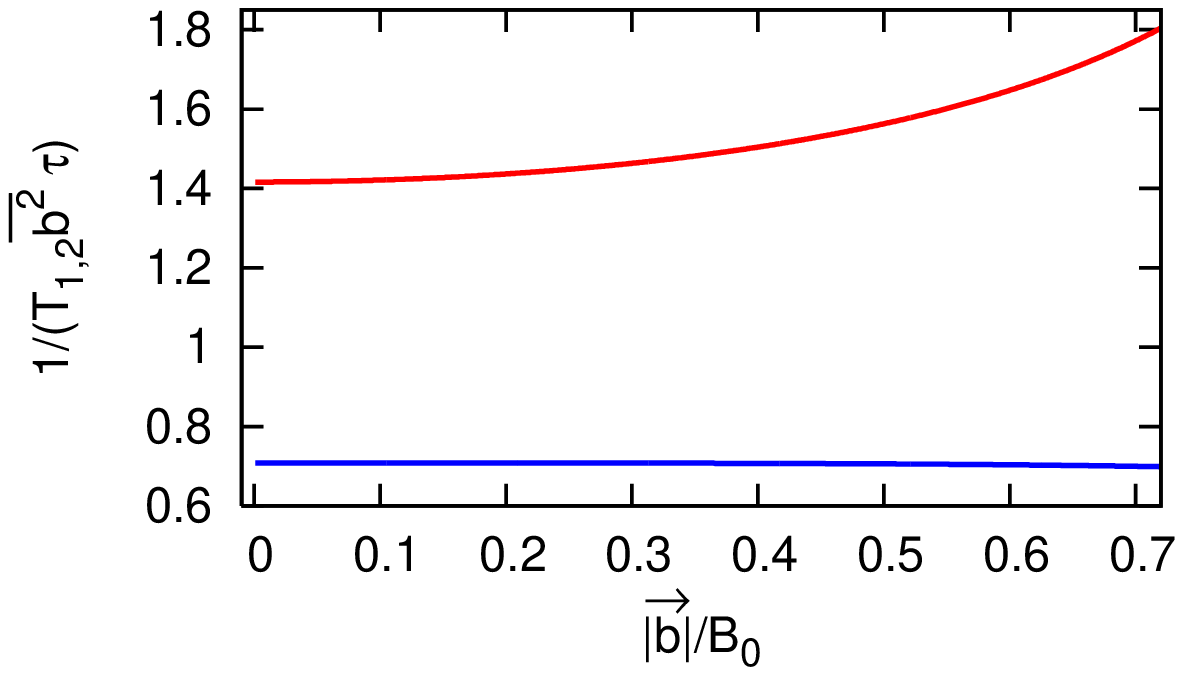}\\ \\
(c)\\
\includegraphics[scale=0.67, angle=-90]{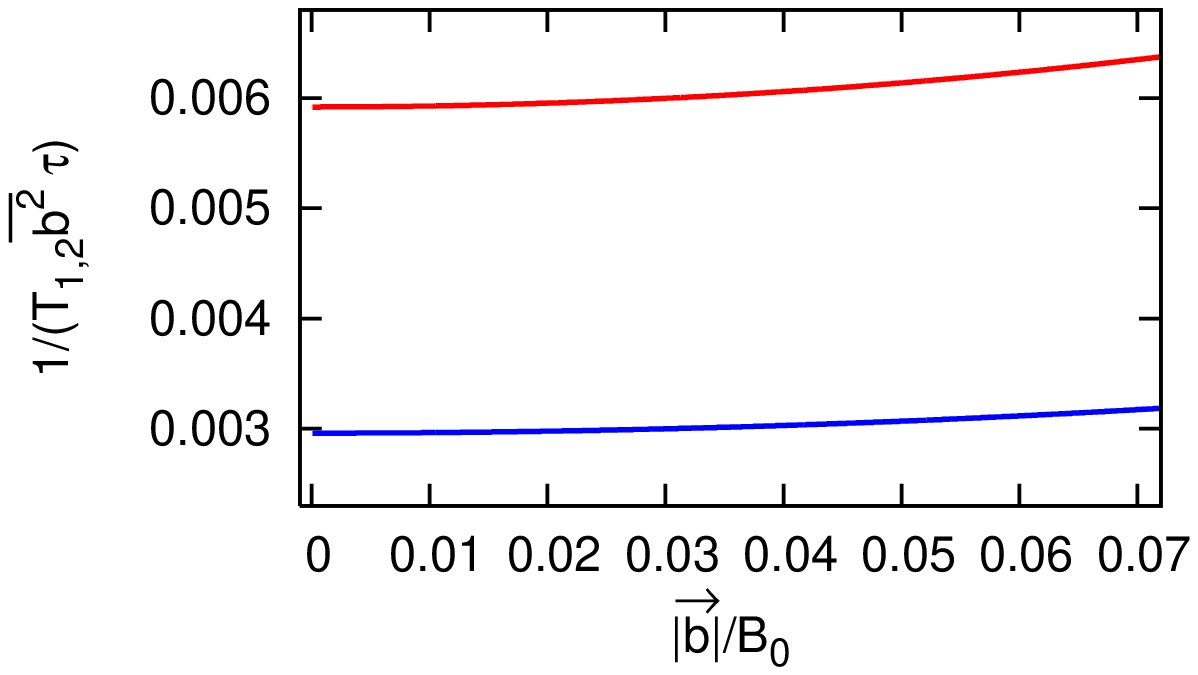}
\end{tabular}
\caption{ $1/(T_{1,2} \overline{b^2}\tau)$ as functions of $\left\vert \vec
{b}\right\vert / B_{0} $
 for $\overline{b_{x}^{2}}$=$\overline{b_{y}%
^{2}}$ and $\overline{b_{z}^{2}}=0$ with $B_{0}\tau=0.1,1$ and 10,
corresponding to (a), (b) and (c) respectively.\ Upper (red) lines are for
$T_{1}$ and lower (blue) lines for $T_{2}$.}%
\end{figure}}

We first look at the case of purely transverse noise: $\overline{b_{x}^{2}%
}=\overline{b_{y}^{2}}$ and $\overline{b_{z}^{2}}=0$. \ The probability
distribution is again the uniform one for the angle $\theta=\tan^{-1}%
(b_{y}/b_{x})$ with $b_{x}^{2}+b_{y}^{2}=b_{0}^{2},$ a constant. \ In Fig. 1
we plot the normalized rates $1/T_{1}\overline{b^{2}}\tau$ and $1/T_{2}%
\overline{b^{2}}\tau$ as a function of $\left\vert \vec{b}\right\vert /B_{0}$
for three values of $B_{0}\tau:B_{0}\tau=0.1,$ $B_{0}\tau=1$ and $B_{0}%
\tau=10.$ \ First note that the correct perturbative limit is obtained when
$\left\vert \vec{b}\right\vert /B_{0}<1$ and $B_{0}\tau<1$ in Fig. 1(a). \ The
standard transverse-noise relation $1/T_{1}=2/T_{2}$ holds over a fairly wide
range: roughly $\left\vert \vec{b}\right\vert /B_{0}\leq3,$ after which
$1/T_{1}>2/T_{2}.$ \ Stronger noise causes $1/T_{1}$ to be a super-quadratic
function of $\overline{b^{2}},$ while the usual physical picture of dephasing
$(1/T_{2})$ as due to a random walk in the phase variable appears to hold for
all values of $\left\vert \vec{b}\right\vert /B_{0}.$ \ As $B_{0}\tau$
increases, Fig. 1(b) and 1(c), we find that the relaxation rates decrease.
\ This is not surprising: as $\tau$ increases, $k_{xx}\left(  \omega\right)
=k_{yy}(\omega)$ become sharply-peaked functions of $\omega$, and the resonant
processes responsible for the relaxation are suppressed. \ As the rates
decrease, the $1/T_{1}=2/T_{2}$ holds over a braoder range of noise strength
$\left\vert \vec{b}\right\vert /B_{0}$. \ 

{\begin{figure}[ptb]
\begin{tabular}{l}
{(a)}\\
\includegraphics[scale=0.67, angle=-90]{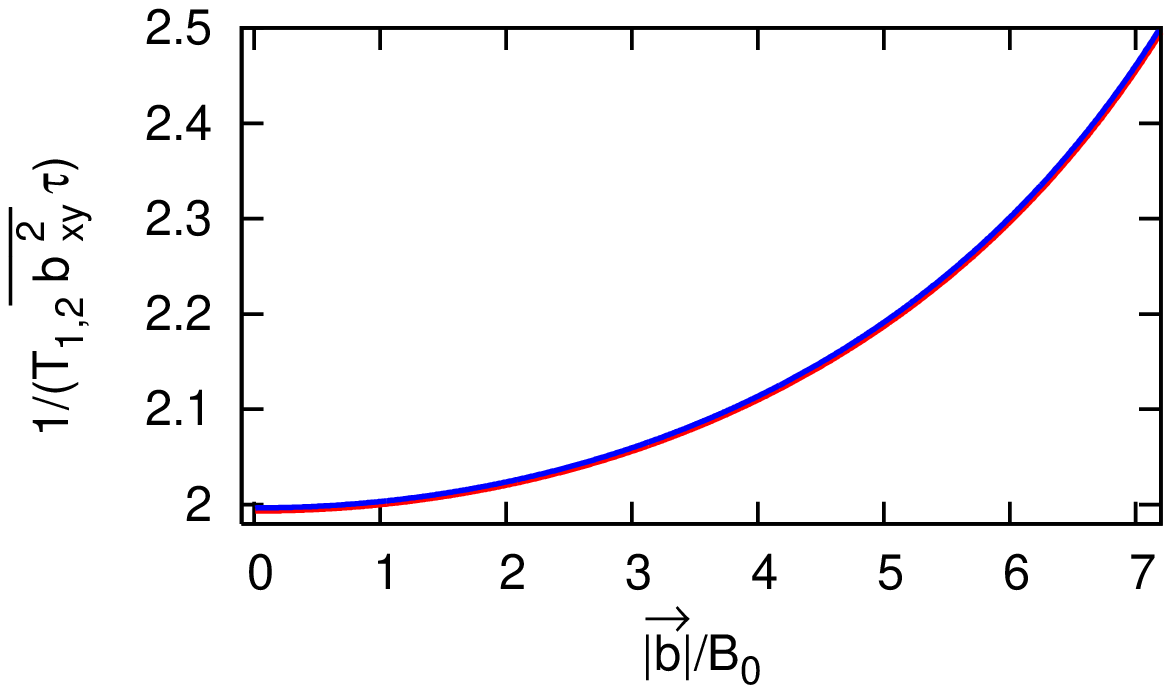}\\ \\
(b)\\
\includegraphics[scale=0.67, angle=-90]{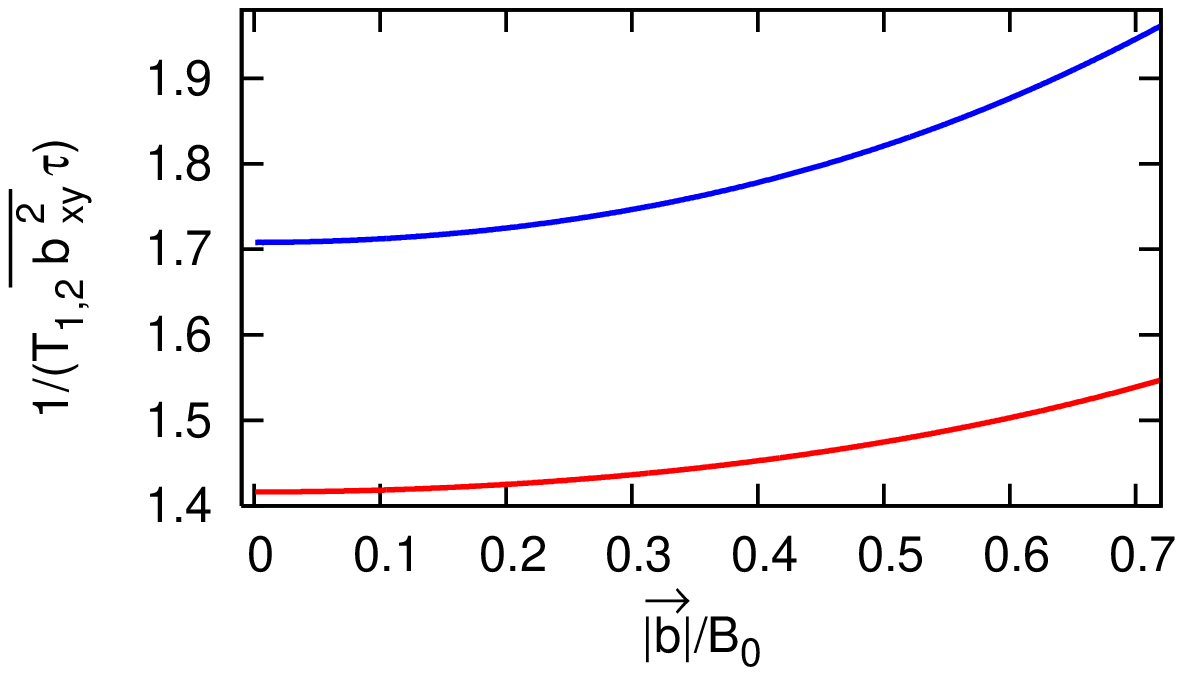}\\ \\
(c)\\
\includegraphics[scale=0.67, angle=-90]{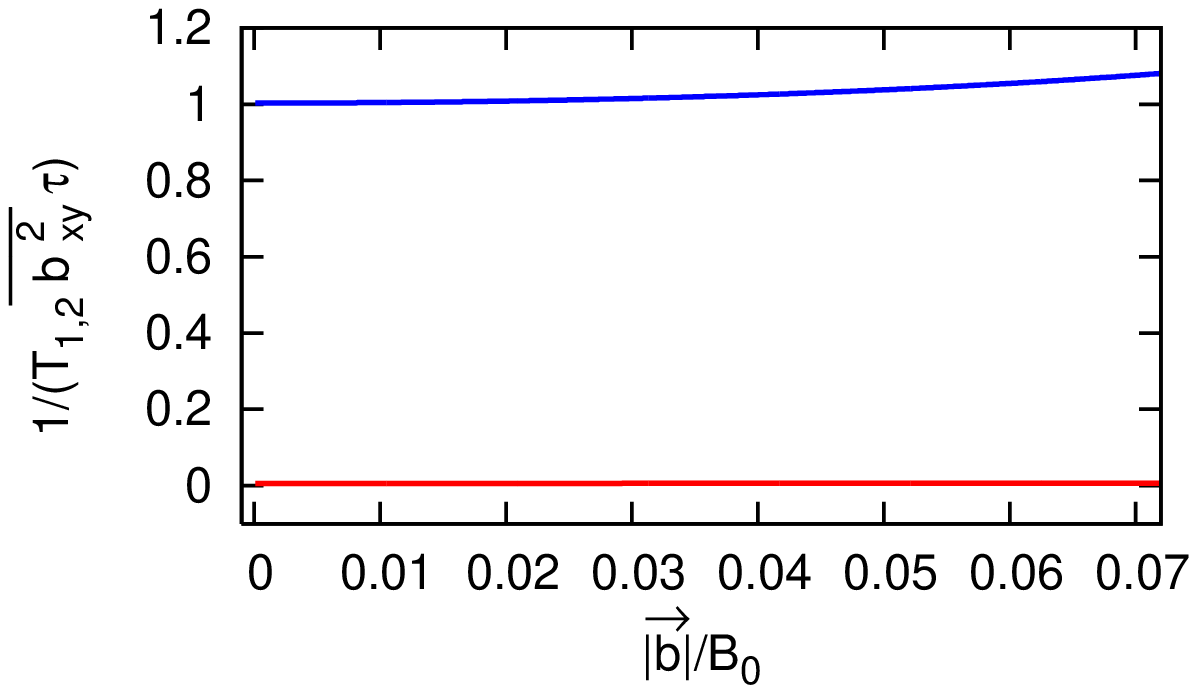}
\end{tabular}
\caption{ $1/(T_{1,2}\overline{b_{xy}^{2}}\tau)$ with 
$\overline{b^2_{xy}}=\overline{b^2_x}+\overline{b^2_y}$
as functions of $\left\vert \vec{b}\right\vert / B_{0} $
 for $\overline{b_{x}^{2}}$=$\overline{b_{y}
^{2}} = \overline{b_{z}^{2}}$ with $B_{0}\tau=0.1,1$ and 10, corresponding to
(a), (b) and (c) respectively.\ Lower (red) lines are for $T_{1}$ and upper
(blue) lines for $T_{2}$. }%
\end{figure}}

Secondly, we consider the case of isotropic noise: $\overline{b_{x}^{2}%
}=\overline{b_{y}^{2}}=\overline{b_{z}^{2}}$ with a uniform distribution of
the solid angle and $b_{x}^{2}+b_{y}^{2}+b_{z}^{2}=b_{0}^{2},$ a constant.\ In
Fig. 2 we plot the normalized rates $1/T_{1}\overline{b^{2}}\tau$ and
$1/T_{2}\overline{b^{2}}\tau$ as a function of $\left\vert \vec{b}\right\vert
/B_{0}$ for the same three values of $B_{0}\tau:B_{0}\tau=0.1,$ $B_{0}\tau=1$
and $B_{0}\tau=10.$ First note that the correct perturbative limit is obtained
when $\left\vert \vec{b}\right\vert /B_{0}<1$ and $B_{0}\tau<1$ in Fig. 2(a):
\ the relation $1/T_{1}=1/T_{2}$ holds quite well for all values of noise
strength $\left\vert \vec{b}\right\vert /B_{0},$ but both $1/T_{1}$ and
$1/T_{2}$ are super-quadratic functions of $\overline{b^{2}}.$ \ As $B_{0}%
\tau$ increases, the dephasing rate $1/T_{2}$ begins to dominate over the
energy relaxation rate $1/T_{1}.$ \ Figs. 2(b) and 2(c), the relaxation rates
decrease for the same reasons as above. \ \ Interestingly, however, the
$1/T_{1}=1/T_{2}$ relation is violated, presumably again due to the varying
relative weights of $k_{xx}(\omega)$ and $k_{zz}(\omega=0)$ as $\tau$ is 
increased.

\section{Correlated-noise models}

\subsection{General solution}

The Hamiltonian remains the same:
\[
H=-B_{0}\sigma_{z}-\vec{b}(t)\cdot\vec{\sigma}=-\vec{B}\cdot\vec{\sigma},
\]
and the piecewise constant time dependence of $\vec{b}(t)$ is retained. \ Now,
however, successive values of $\vec{b}_{i}$ may be correlated. \ Thus we take
the following model for $P(\vec{b}_{1},\vec{b}_{2},..,\vec{b}_{m})$ : \ the
values of $\vec{b}_{i}$ are independent except for neighboring intervals, so
we have
\begin{equation*}
P(\vec{b}_{1},\vec{b}_{2},..,\vec{b}_{m})=\Pi_{i=1,m=1}P(\vec{b}_{i},\vec
{b}_{i+1})
\end{equation*}
\ We shall make the natural assumption that $P(\vec{b},\vec{b}^{\prime})$ is
symmetric: $P(\vec{b},\vec{b}^{\prime})=P(\vec{b}^{\prime},\vec{b}).$

Now $\vec{\sigma}_{ij}$ transforms in the first interval into%
\begin{align}
\overline{\vec{\sigma}_{ij}(\tau)}(\vec{b}_{2})=&\int d\vec{b}_{1}%
P(\vec{b}_{1},\vec{b}_{2})\left[  U_{1}^{+}(\tau)\right]  _{ik}|\left[
U_{1}(\tau)\right]  _{lj}\vec{\sigma}_{kl}\nonumber\\
\,=&\int d\vec{b}_{1}P(\vec{b}_{1},\vec{b}_{2})\left[  \exp iH_{1}\tau\right]
_{ik}\left[  \exp-i\vec{H}_{1}\tau\right]  _{lj}\vec{\sigma}_{kl}\nonumber\\
=&\int d\vec{b}_{1}P(\vec{b}_{1},\vec{b}_{2}) \nonumber\\
& \times \left[E\cos(B_1\tau)
-i\widehat{B}_{1}\cdot\vec{\sigma}\sin(B_{1}\tau)\right]  _{ik}
\nonumber\\
& \times \left[
E\cos(B_{1}\tau)+i\widehat{B}_{1}\cdot\vec{\sigma}\sin(B_{1}\tau)\right]
_{lj}\vec{\sigma}_{kl},
\end{align}
where $E$ is the identity matrix and summation over repeated indices is
implied. \ In matrix form, this is%
\begin{align}
\overline{\vec{\sigma}(\tau)}(\vec{b}_{2})=&I_{0}(\vec{b}_{2})~\vec{\sigma
}+\sum_{ij}I_{ij}(\vec{b}_{2})~\sigma_{i}\vec{\sigma}\sigma_{j}+\nonumber\\
&+i\sum_{i}%
I_{i}(\vec{b}_{2})~\left(  \vec{\sigma}\sigma_{i}-\sigma_{i}\vec{\sigma
}\right)  ,
\end{align}
where the integrals $I_{0},~I_{i}$ and $I_{ij}$ are defined as%
\begin{align*}
I_{0}(\vec{b}^{\prime})  &  =\int P(\vec{b},\vec{b}^{\prime})\cos^{2}%
(B\tau)~d\vec{b}\\
I_{i}(\vec{b}^{\prime})  &  =\int P(\vec{b},\vec{b}^{\prime})\widehat{B}%
_{i}\sin(B\tau)\cos\left(  B\tau\right)  ~d\vec{b}\\
I_{ij}(\vec{b}^{\prime})  &  =\int P(\vec{b},\vec{b}^{\prime})\widehat{B}%
_{i}\widehat{B}_{j}\sin^{2}(B\tau)~d\vec{b},
\end{align*}
analogously to the previous case. \ 

We may write the evolution equations in a compact fashion by defining
$f_{0}(\vec{b})=\cos^{2}(B\tau),~f_{i}(\vec{b})=\widehat{B}_{i}\sin(B\tau
)\cos\left(  B\tau\right)  ,~$and $f_{ij}(\vec{b}^{\prime})=\widehat{B}%
_{i}\widehat{B}_{j}\sin^{2}(B\tau),$ and $T_{xx}(\vec{b})=f_{0}(b)+f_{xx}%
(\vec{b})-f_{yy}(\vec{b})-f_{zz}(\vec{b}),$ etc., as in Eq.\ \ref{eq:matrix}.
\ Then we find
\[
\vec{\sigma}(\vec{b}_{2},\tau)=\int d\vec{b}_{1}P(\vec{b}_{2},\vec{b}%
_{1})T(\vec{b}_{1})~\vec{\sigma}(0),
\]
where $\vec{\sigma}$ is to be interpreted as a column matrix. \ When this
process is iterated we have%
\begin{align*}
\vec{\sigma}(\vec{b}_{3},2\tau) =& \int d\vec{b}_{2}P(\vec{b}_{3},\vec
{b}_{2})T(\vec{b}_{2})~\vec{\sigma}(\vec{b}_{2},\tau)\\
  =&\int d\vec{b}_{2}P(\vec{b}_{3},\vec{b}_{2})T(\vec{b}_{2})\\
 &\times \int d\vec{b}_1 P(\vec{b}_2,\vec{b}_1)~T(\vec{b}_{1})~\vec{\sigma}(0),
\end{align*}
and finally%
\begin{align*}
\vec{\sigma}(m\tau)=&\int d\vec{b}_{m+1}\int d\vec{b}_{m}P(\vec{b}_{m+1}%
,\vec{b}_{m})T(\vec{b}_{m})\cdot\cdot\cdot\times \\
&\int d\vec{b}_{2}P(\vec{b}_{3}%
,\vec{b}_{2})T(\vec{b}_{2})\int d\vec{b}_{1}P(\vec{b}_{2},\vec{b}_{1}%
)~T(\vec{b}_{1})~\vec{\sigma}(0).
\end{align*}
This problem is exactly soluble when the function $P(\vec{b},\vec{b}^{\prime
})$ is separable:%

\begin{equation}
P(\vec{b},\vec{b}^{\prime})=\sum_{n=1}^{N}p_{n}(\vec{b})p_{n}(\vec{b}^{\prime
}).
\end{equation}
This is of course not true for all functions $P(\vec{b},\vec{b}^{\prime})$.
\ However, if $P(\vec{b},\vec{b}^{\prime})$ is continuous, it can be
approximated uniformly by such an expression \cite{courant}. \ Substituting
and converting to index notation with repeated indices summed, we have%
\begin{align*}
\sigma_{i}(m\tau)= & \int d\vec{b}_{m+1}\int d\vec{b}_{m}p_{n_{m}}(\vec{b}%
_{m+1})p_{n_{m}}(\vec{b}_{m})T_{ii_m}(\vec{b}_{m})\cdot\cdot\cdot\\
 &\times 
 \int d\vec{b}_{2}p_{n_{2}}(\vec{b}_{3})p_{n_{2}}(\vec{b}_{2})T_{i_{3}i_{2}}(\vec
{b}_{2}) \\
 &\times \int d\vec{b}_{1}p_{n_{1}}(\vec{b}_{2})p_{n_{1}}(\vec{b}_{1}%
)~T_{i_{2}j}(\vec{b}_{1})~\sigma_{j}(0).
\end{align*}
This can now be written as a matrix product%
\begin{widetext}
\begin{equation}
\sigma_{i}(m\tau)=\int d\vec{b}_{m+1}p_{n_{m}}(\vec{b}_{m+1})S_{\left(
n_{m},i\right)  ,\left(  n_{m-1},i_{m}\right)  }S_{\left(  n_{m-1}%
,i_{m}\right)  ,\left(  n_{m-2},i_{m-1}\right)  }\cdot\cdot\cdot S_{\left(
n_{2},i_{3}\right)  ,\left(  n_{_{1}},i_{2}\right)  }\int d\vec{b}_{1}%
p_{n_{1}}(\vec{b}_{1})~T_{i_{2}j}(\vec{b}_{1})~\sigma_{j}(0),
\end{equation}
\end{widetext}
where we have defined the $3N\times3N$ matrix%
\begin{equation}
S_{\left(  n^{\prime},i\right)  ,\left(  n^{\prime\prime},j\right)  }=\int
d\vec{b}~p_{n^{\prime}}(\vec{b})T_{ij}(\vec{b})p_{n^{\prime\prime}}(\vec{b}).
\end{equation}
Diagonalizing $S,$ we have the solution%
\begin{align}
\sigma_{i}(m\tau)=&\int d\vec{b}^{\prime}F_{n}(\vec{b}^{\prime})~\left(
S^{m-1}\right)  _{\left(  n,i\right)  ,\left(  n^{\prime},j\right) }\times
\nonumber\\
\,& \,\,\, \int
d\vec{b}~F_{n^{\prime}}(\vec{b})T_{jk}(\vec{b})~\sigma_{k}(0)\nonumber \\
\,=& \int d\vec{b}^{\prime}F_{n}(\vec{b}^{\prime})\left\{  R_{S}^{-1}\left[
diag(S)^{m-1}\right]  R_{S}\right\}  _{\left(  n,i\right)  ,\left(  n^{\prime
},j\right)  } \nonumber\\
\,& \,\,\,\times \int d\vec{b}~F_{n^{\prime}}(\vec{b})T_{jk}(\vec{b})~\sigma
_{k}(0).
\end{align}
Here $R_{S}$ is the matrix whose rows are the eigenvectors of $S,$ and
$diag(S)$ is the diagonal matrix with the eigenvalues of $S$ along the
diagonal. The relaxation times are then obtained by the same arguments as in
the previous section, except now they must be obtained by computing the
eigenvalues of $S,$ not $T.$ \ \ The matrices enclosing \ $diag(S)^{m-1}$
determine how to prepare and measure the system if one wishes to determine
some particular relaxation time. \ 

\bigskip

\subsection{Example}

In the first example, the scattering was purely s-wave: isotropic in $\phi$.
\ If the scattering can also occur in the p-wave channel, then the successive
values of $\vec{b}$ are correlated. \ An interesting example is a mixture of
s-wave and p-wave given by%

\begin{equation}
P(\vec{b},\vec{b}^{\prime})=\frac{1}{2\pi b_{0}}\delta(b-b_{0})\delta
(b_{z})\left[  1+r\cos\left(  \phi-\phi^{\prime}\right)  \right]  ,
\end{equation}
with $0\leq r\leq1.$ \ We choose this model partly because of its simplicity.
\ However, it also has some interesting physics. \ For $r=0$ it reduces to the
previous example. \ For $r=1,$ there is suppression of backscattering at
$\phi-\phi^{\prime}=\pi$. \ This implies that, as $r$ increases, the problem
crosses over from a purely random walk in spin space to one in which the
successive steps tend to be in the same direction. \ 

After performing the integrals over the $\left\vert \vec{b}_{i}\right\vert $
and $b_{iz},$ we find%
\begin{align}
P(\phi,\phi^{\prime})  &  =\frac{1}{2\pi}\left[  1+r\cos\left(  \phi
-\phi^{\prime}\right)  \right] \nonumber\\
&  =\frac{1}{2\pi}\left(  1+r\cos\phi\cos\phi^{\prime}+r\sin\phi\sin
\phi^{\prime}\right)
\end{align}
so that the function is separable
\begin{equation}
P(\phi,\phi^{\prime})=\sum_{n=1}^{3}p_{n}(\phi)p_{n}(\phi^{\prime})
\end{equation}
with
\begin{equation}
p_{1}(\phi)=\sqrt{\frac{1}{2\pi}},~p_{2}(\phi)=\sqrt{\frac{r}{2\pi}}\cos
\phi,~p_{3}(\phi)=\sqrt{\frac{r}{2\pi}}\sin\phi.
\end{equation}
We now need to compute the $9\times9$ matrix%
\begin{equation}
S_{(ni),(n^{\prime}j)}=\int d\phi~p_{n}(\phi)T_{ij}(\phi)p_{n^{\prime}}(\phi).
\end{equation}
We are interested here in the decoherence limit: $b_{0}\tau<<B_{0}\tau<<1$.
\ The integrations over the angular functions are elementary. \ The nonzero
elements of $S$ are%
\begin{align*}
S_{(1x),(1x)}  &  =S_{(1y),\left(  1y\right)  }=1-2B_{0}%
^{2}\tau^{2}-b_{0}^{2}\tau^{2}\\
S_{(2x),(2x)}  &  =S_{(3y),\left(  3y\right)  }=\left(
r/2\right)  \left(  1-2B_{0}^{2}\tau^{2}\right)  -rb_{0}^{2}\tau^{2}/2\\
S_{(3x),(3x) }  &  =S_{(2y),\left(  2y\right)  }=\left(
r/2\right)  \left(  1-2B_{0}^{2}\tau^{2}\right)  -3rb_{0}^{2}\tau^{2}/4\\
S_{(1z),(1z)}  &  =1-2b_{0}^{2}\tau^{2}\\
S_{(2z),(2z)}  &  =S_{(3z),\left(  3z\right)  }=\left(
r/2\right)  \left(  1-2b_{0}^{2}\tau^{2}\right) \\
S_{(1x),(1y)}  &  =-S_{\left(  1y\right)  ,\left(  1x\right)
}=2B_{0}\tau\\
S_{(2x),(2y)} & =S_{(3x),(3y)} =-S_{(2y),(2x)}=-S_{(3y),(3x)}=rB_{0}\tau\\
S_{(2x),(3y)}  &  =S_{(2x),\left(  3y\right)  }=S_{(2y),(3x)}
=S_{(2y),(3x)}=rb_{0}^{2}\tau^{2}/2\\
S_{(1x),(2z)}  &  =S_{(2x),\left(  1z\right)  }=S_{(1z),\left(
2x\right)  }=S_{(2z),(1x)}\\
  & =S_{(1y),(3z)}
=S_{(3y),(1z)} =S_{(1z),(3y)} \\
  &=S_{(3z),(1y)}=\sqrt{r} b_0 B_{0}\tau^{2}/2\\
S_{(1x),(3z)} & =S_{(3x),(1z)}=-S_{(1y),(2z)}=-S_{(2y),(1z)}\\
  &=-S_{(1z),(3x)}=-S_{(3z),(1x)} =S_{(1z),( 2y)}\\
  &=S_{(2z),(1y)}=-\sqrt{r}b_{0}\tau/2
\end{align*}
This matrix is easily diagonalized numerically, and the results are shown in
Fig. 3. \ All eigenvalues are less than one in magnitude, as before. \ Six of
these eigenvalues are much less than 1, and correspond to transients. \ Two
(one) give finite relaxation times when $t_{f}/\tau$ becomes large and
correspond to $T_{2}$ ($T_{1}).$ $\ 1/T_{1}$ increases from $2b_{0}^{2}\tau$
to about $5.9~b_{0}^{2}\tau,$ and $1/T_{2}$ increases from b$_{0}^{2}\tau$ to
$2.5$~$b_{0}^{2}\tau$ as $x$ goes from $0$ to $1.$ \ This is to be expected,
as decrease in the average angle of change of the field corresponds
intuitively to a longer step in the random walk of the spin vector.

\begin{figure}[ptb]
\includegraphics[scale=.67, angle=-90]{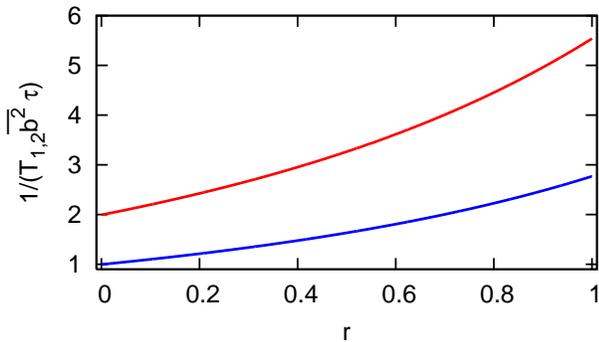}
\caption{$1/(T_{1,2}\overline{b^2}\tau)$ as functions of
$r$. The upper (red) line is for $T_{1}$ and the lower (blue) one for $T_{2}%
$.}%
\end{figure}

\section{Discussion}

The two basic features that make the basic white-noise-like model solvable is
that (1) the noise Hamiltonian is piecewise constant with equal intervals and
that (2) the system space is two-dimensional. \ These features limit the
applicability of the model. \ 

Condition (1) is a fairly severe limitation:\ often, the effectiveness of the
noise is related to $k_{ii}(\omega_{0}).$ \ In the basic model, this can only
be varied by changing the parameter $B_{0}\tau.$ \ The \textit{shape} of
$k_{ii}(\omega_{0})$ is fixed in the white noise model, and can only be
changed to a limited extent by going over to correlated-noise generalizations.
\ Some very interesting problems, such as the effect of $1/f$ noise on qubits
\cite{schon}, appear to be outside the scope of this method. \ It may be
possible to allow for a distribution of intervals without sacrificing
solvability - the multiple time integral thus obtained, though complicated,
may be susceptible to integral transform techniques.

Condition (2) is not easily jettisoned. \ The three Pauli matrices generate
the 3-dimensional algebra of $SU(2),$ $T$ generates orbits in this algebra,
and this implies that $T$ is a $3\times3$ matrix. \ The transformations of a
three-dimensional Hilbert space (for a qtrit) belong to $SU(3),$ and the
algebra is an 8-dimensional space. \ Hence the transfer $T$ matrix for this
problem would be $8\times8.$ \ There may be special situations in which there
is an exact solution. \ The solution of the two-dimensional Ising model
depends on diagonalizing a $2^{N}\times2^{N}$ transfer matrix, where $N$ is
the number of spins in the transverse direction. \ The corresponding
generalization of the present problem would require the diagonalization of an
$N^{2}-1\times N^{2}-1$ matrix for an $N$-level system. \ $N^{2}-1$ is the
number of generators of $SU(N).$ \ Such a generalization is likely to require
some new ideas.

Despite these limitations, the method is new and susceptible to development in
different directions. \ It should serve as an important tool in the study of
decoherence in the future.

We would like to acknowledge useful converstaions with C.\ Tahan and M.
Friesen, and the support of the NSF ITR program.

\section{Appendix}

Here we show that the eigenvalues satisfy $\left\vert d_{3}\right\vert \leq1$
and $\left\vert d_{\pm}\right\vert \leq1.$

Theorem: $\left\vert d_{\pm}\right\vert \leq1.$

\bigskip Proof:%
\begin{align*}
\left\vert d_{\pm}\right\vert ^{2} =&\left\vert I_{0}-I_{zz}\pm
i\sqrt{4I_{z}^{2}-\left(  I_{xx}-I_{yy}\right)  ^{2}}\right\vert ^{2}\\
 = & \left(  I_{0}-I_{zz}\right)  ^{2}+4I_{z}^{2}-\left(  I_{xx}-I_{yy}\right)
^{2}\\
&  \leq\left(  I_{0}-I_{zz}\right)  ^{2}+4I_{z}^{2}\\
= & \left\{  \int P(\vec{b})\left[  \cos^{2}(B\tau)-\widehat{B}_{z}\sin
^{2}(B\tau)\right]  d^{3}b\right\}  ^{2} + \\
\,& \,\, +4\left[  \int P(\vec{b})\widehat
{B}_{z}\sin(B\tau)\cos\left(  B\tau\right)  d^{3}b\right]  ^{2}\\
\leq & \int\left[  \cos^{2}(B\tau)-\widehat{B}_{z}\sin^{2}(B\tau)\right]
^{2}P(\vec{b})d^{3}b + \\
\, & + 4\int\left[  \widehat{B}_{z}\sin(B\tau)\cos\left(
B\tau\right)  \right]  ^{2}P(\vec{b})d^{3}b,
\end{align*}
where the final inequality is an application of the Schwarz inequality.
\ Since $\left\vert \widehat{B}_{z}\right\vert \leq1,$ we also have
\begin{align*}
\left\vert d_{\pm}\right\vert ^{2} \leq& \int\left[  \cos^{2}(B\tau
)-\widehat{B}_{z}\sin(B\tau)\right] ^{2}P(\vec{b})~d^{3}b+\\
&+4\int\left[
\widehat{B}_{z}\sin(B\tau)\cos\left(  B\tau\right)  \right]  ^{2}P(\vec
{b})~d^{3}b\\
  =&\int\left[  \cos^{2}(B\tau)+\widehat{B}_{z}\sin^{2}(B\tau)\right]
^{2}P(\vec{b})~d^{3}b\\
  \leq&\int\left[  \cos^{2}(B\tau)+\sin^{2}(B\tau)\right]  ^{2}P(\vec
{b})~d^{3}b\\
 =&\int P(\vec{b})~d^{3}b\\
 =&1.
\end{align*}
Theorem: $\left\vert d_{3}\right\vert \leq1.$

Proof:

\begin{align*}
d_{3}  &  =I_{0}-I_{xx}-I_{yy}+I_{zz}\\
&  =\int P(\vec{b})\left[ \cos^{2}(B\tau)+\left( \widehat{B}_z
^{2}-\widehat{B}_x^2-\widehat{B}_y^2 \right) \sin^2(B\tau)\right]
d^{3}b\\
&\leq\int P(\vec{b})\left[\cos^{2}(B\tau)+
        \widehat{B}_z^2\sin ^2(B\tau)\right] d^3 b\\
&  \leq\int P(\vec{b})\left[ \cos^2 (B\tau)+\sin^2 (B\tau)\right] d^3 b\\
&=1,
\end{align*}
and recalling that $\widehat{B}_{z}^{2}+\widehat{B}_{x}^{2}+\widehat{B}%
_{y}^{2}=1,$ we also have
\begin{align*}
d_{3}  &  =\int P(\vec{b})\left[  \cos^{2}(B\tau)+\left(  \widehat{B}_{z}%
^{2}-\widehat{B}_{x}^{2}-\widehat{B}_{y}^{2}\right)  \sin^{2}(B\tau)\right]
d^{3}b\\
&  =\int P(\vec{b})\left\{  \cos^{2}(B\tau)+\left[  1-2\left(  \widehat{B}%
_{x}^{2}+\widehat{B}_{y}^{2}\right)  \right]  \sin^{2}(B\tau)\right\}
d^{3}b\\
&  =1-2\int P(\vec{b})\left[  \cos^{2}(B\tau)+\sin^{2}(B\tau)\left(
\widehat{B}_{x}^{2}+\widehat{B}_{y}^{2}\right)  \right]  d^{3}b\\
&  \geq1-2\int P(\vec{b})\left[  \cos^{2}(B\tau)+\sin^{2}(B\tau)\right]
d^{3}b\\
&  =-1.
\end{align*}
Hence $\left\vert d_{3}\right\vert \leq1.$

\end{document}